\documentclass[12pt,letterpaper]{article} \usepackage[latin1] {inputenc} 
\usepackage{amsmath} \usepackage{amsfonts} \usepackage{amssymb} 
\usepackage{amscd}
\begin{document}
\title{Equations of motion, Noncommutativity and Quantization}
\author{Ignacio Cortese\footnote{nachoc@nucleares.unam.mx}\,\, and  J. Antonio Garc\'\i a\footnote{ garcia@nucleares.unam.mx}\\
\em Instituto de Ciencias Nucleares, \\
\em Univesidad Nacional Aut\'onoma de M\'exico\\
\em Apartado Postal 70-543, M\'exico D.F., M\'exico}
\maketitle
\abstract{We study the relation between a given set of equations of motion in configuration space and a Poisson bracket. A Poisson structure is consistent with the equations of motion if the symplectic form satisfy some consistency conditions. When the symplectic structure is commutative these conditions are the Helmholtz integrability equations for the nonrestricted inverse problem of the calculus of variations \cite{Hojman-Shepley}. We have found the corresponding consistency conditions for the symplectic noncommutative case.}

\section{Introduction}

A deep reconsideration of some fundamental concepts of the past 
century physics is a sign of our present time.  One of these fundamental concepts is space-time that was largely considered as a background scenario where the dynamics take place.      
In addition, it is now believed that some intrinsic properties of space-time could be very different from the usual ones at some high energy scale. 
Inspired from Open String Theory and Loop Quantum Gravity the idea that the smooth geometry of space-time has to be replaced with a noncommutative geometry 
\cite{Douglas} is now a very active area in the quest to search for new physics whose wide scope runs from phenomenology to physical theory and mathematical research. In string theory this idea can  be implemented from a certain sigma model with appropriate boundary conditions. When we allow for space-space noncommutativity the low energy limit of this model produces a 
 noncommutative field
theory\cite{SW}. If we allow for space-time noncommutativity a particular low energy limit produces a noncommutative open string theory
\cite{SST}. The noncommutative field theories are
defined by a deformation of the algebra of classical field functions
on space-time where the standard product is  replaced with a new noncommutative 
$\star$ Weyl-Moyal product \cite{Douglas-Nekrasov}. Generalizations of these ideas to the case when the noncommutativity parameter depends on space-time are also a current area of research \cite{Konsev, Cattaneo-Felder}. 

These theories are deformations of classical field theories in the same sense as a particular deformation of a classical theory can produce a quantum theory. 
In order to include the effects of noncommutativity in classical and quantum dynamics
we could wonder if this deformation procedure can also be applied in such cases.
The answer is in the affirmative but the structure to be deformed is not the space time but the phase-space symplectic structure. A clear objection to such procedure is that a symplectic structure can be deformed by a redefinition of variables in phase space (Darboux transformations). We can always recast the symplectic structure in the standard Darboux form using the $(q,p)$ phase space variables. As a result of this transformation we will end with a new  modified Hamiltonian. It is indeed possible to construct interacting models from a deformation of $\sigma$. In general these interactions do not preserve symmetries and are non-local upon quantization. 
The standard lore is to define a noncommutative classical system (NCCS) by a Hamiltonian of the form $H=T+V$ and a symplectic structure that implements the noncommutativity between the coordinates of the configuration space. This two classically equivalent perspectives are, in general, not equivalent at quantum level (the Darboux transformation is {\it not} a unitary transformation in quantum mechanics). 
 
This approach can be implemented also in field theory \cite{Gamboa}. We will call these resulting field theories {\it symplectic noncommutative field theories}. They include the effects of  noncommutativity through a deformation of the symplectic structure with the property that the Poisson bracket between fields at different space points is different from zero. The new parameter that controls this symplectic noncommutativity could be used as a natural cut off  and/or as a model for Lorentz violation at certain scale.
 
It is well know that we can specify the dynamical content of a classical theory in different but classically equivalent ways: a) by the equations of motion in configuration space (i.e., the forces), b) by a Lagrangian in the tangent space of the manifold that define the configuration space or c) by a symplectic structure $\sigma$ and a Hamiltonian\footnote{The relation between a first order formulation of the dynamics and the inverse problem of the calculus of variations in tangent space was worked out in \cite{Henneaux} and \cite{CG}. The inverse problem of the calculus of variations  in the first order formulation was studied in \cite{HU}.}. Quantum mechanics needs global information from the classical mechanics description. This global information is used by quantum theory to account for fluctuations of classical solutions\footnote{Recall for example the path integral approach to quantization.}. Indeed the information encoded in a) is not sufficient to quantize a classical system. For NCCS the natural way to specify the dynamical information is to encode it in the form c). 
Here we are interested in 
the consequences of symplectic noncommutativity in {\em configuration space}.  
We will provide the dynamical information by means of the equations of motion {\it and} the symplectic structure. This information can not be specified in an arbitrary way. Given the forces, the symplectic structure $\sigma$ must satisfy some dynamical consistency conditions. 
The aim of this letter is to find and analyze these dynamical consistency conditions between the given forces and a noncommutative symplectic structure. 
 Generalizing the analysis presented in \cite{Hojman-Shepley} where the authors found that the dynamical consistency conditions between commutative commutation relations (CCR) and a given set of equations of motion in configuration space are precisely the Helmholtz conditions for the existence of a Lagrangian in tangent space, we will present here the corresponding dynamical consistency conditions for the case of
noncommutative commutation relations (NCR). We will show  that these consistency conditions do not imply the existence of a Lagrangian for the given set of equations of motion ({\em i.e.}, they are not any more the Helmholtz conditions). They could be regarded as a deformation of the Helmholtz conditions in the sense that they can be written as the Helmholtz conditions plus terms that depend on the noncommutative parameter $\theta$. If one insist on the existence of a Lagrangian in tangent space the Helmholtz conditions will be satisfied. It turn out that the remaining conditions on the forces that comes from these $\theta$ dependent terms are very restrictive. In fact, for isotropic central potentials only polynomials up to second degree in the coordinates satisfy these conditions.

The quantization of a given set of equations of motion in configuration space can be performed either by using CCR or NCR providing that the corresponding consistency conditions are satisfied. The resulting quantum theories are in general nonequivalent. 
It also could be the case that the system can not be quantized
using CCR because it does not have a Lagrangian ({\it i.e.}, the standard consistency conditions are not satisfied) but nevertheless it can be quantized in a consistent way using NCR (when our new consistency conditions are satisfied).
As a consequence of our results we can state that a consistent quantization of the given equations of motion is possible by solving the new conditions, even when a Lagrangian function does not exist. 

We will present also a generic solution to these new consistency conditions. This solution is the analogous of the standard solution of the inverse problem of the calculus of variations for forces that come from a potential function.

Previous attempts to analyze the same problem that we are addressing  here are \cite{Acatrinei, Sedra}, but the scope of these references is very limited and in the case of the second reference the proposed generalization of the CCR is inconsistent. That noncommutative symplectic structures may have deep consequences in the study of the inverse problem of the calculus of variations in tangent space was first observed in \cite{Hojman, Novikov, Henneaux}. In particular, in \cite{Hojman} the author conclude that the equations of motion in configuration space should be deformed in such a way that they are not more of second order in time derivatives but of third order. We are exploring the possible relation between this result and the consistency conditions that we have found here.

The organization of the letter is as follows: in section 2 we will present our new consistency conditions generalizing previous results. In section 3 a generic solution to the new conditions is constructed. We present an nonequivalent quantization for the case of the harmonic oscillator and a consistent quantization of  a system that does not have a Lagrangian. Finally,  the conclusions and some possible consequences of the extension of our results to the case of field theory are the content of section 4.

\section{Consistency conditions for NCR}

Given a set of equations of motion in configuration space 
\begin{equation}
\label{eq-m}
\ddot x^i - F^i(x^j,\dot x^j,t)=0,
\end{equation}
and the NCR\footnote{ 
We will identify the commutation relations with the corresponding classical Poisson brackets relations. At a formal level we can use canonical quantization or Moyal-Wigner deformation quantization to relate our results with the quantum ones.
If $\sigma$ depends explicitly on time we can reparametrize the classical theory and use the technics developed in \cite{GVU}   to obtain results analogous to the ones presented here.} 
\begin{equation}
\label{NCR}
[x^i,\dot x^j]=g^{ij}(x,\dot x),\qquad [x^i,x^j]=\theta^{ij},
\end{equation}
where $g^{ij}$ is a symmetric matrix and $\theta^{ij}$ an antisymmetric constant matrix, we say that the NCR are compatible with the given equations of motion (\ref{eq-m}) if using\footnote{Notice that this condition is not the standard Leibnitz rule. In fact it implies a dynamical compatibility between the bracket and the dynamical vector field along the solution curves of the equations of motion. If we define the bracket 
by
$$[A,B]=\frac{\partial A}{\partial z^a}\sigma^{ab}\frac{\partial B}{\partial z^b},$$
where $z^a=(x^i,\dot x^j)$ then the condition (\ref{leibniz}) imply
$${\cal L}_F(\sigma^{ab})=0,$$
where ${\cal L}_F$ is the Lie derivative along the solution vector field associated with the system (\ref{eq-m}). The content of this condition will appear as the Helmholtz conditions in the following paragraphs.}
\begin{equation}
\label{leibniz}
{\frac{D}{Dt}}[A,B]=[{\frac{D}{Dt}}A,B] + [A,\frac{D}{Dt}B],
\end{equation}
where
$${\frac{D}{Dt}}=F^i\frac{\partial}{\partial\dot x ^i}+\dot x^i\frac{\partial}{\partial x ^i}+\frac{\partial}{\partial t},$$
we can construct a matrix $g^{ij}$ and a matrix $B^{ij}$ defined by 
$$B^{ij}=[\dot x^i,\dot x^j],$$ 
that depends on $g^{ij},\theta^{ij}$ and $F^i$ in such a way
that the corresponding symplectic matrix $\sigma$ is consistent with the Leibnitz rule (\ref{leibniz}) and the Jacobi identity. So we will have a set of conditions that involves the basic matrices $g^{ij},\theta^{ij}$ and the forces that define the system $F^i$. We will call these conditions {\em dynamical consistency conditions}. When the noncommutative parameter $\theta^{ij}$ is zero, it turns out that these consistency conditions are the Helmholtz conditions for the non restricted inverse problem of the calculus of variations \cite{Anderson} associated with the equations of motion (\ref{eq-m}) {\em i.e.,} the integrability conditions of the partial differential equations  
$$EL_i(L)=g_{ij}(\ddot x^j - F^j(x,\dot x,t)),$$
for the function $L$. Here
$EL_i$ is the Euler-Lagrange operator and $L$ is a Lagrangian function for the given system. Indeed when a Lagrangian exists, we can conclude following \cite{Hojman-Shepley}, that a consistent quantization of the system is possible if and only if it comes from a variational principle with Lagrangian $L$. This is the content of the No Lagrangian?, No quantization! lema.

Our task is now to find these consistency conditions when $\theta^{ij}$ is different from zero. A straightforward calculation gives
\begin{subequations}
\label{helm}
\begin{align}
[x^i,g^{jk}]=[x^j,g^{ik}],\label{h-a}\\
\frac{D}{Dt}g^{ij}=\frac12 [x^i,F^j]+\frac12 [x^j,F^i],\label{h-b}\\
\frac{D}{Dt}B^{ij}= [\dot x^i,F^j]- [\dot x^j,F^i],\label{h-c}
\end{align}
\end{subequations}
where 
\begin{multline}
\label{B}
B^{ij}=[\dot x^i,\dot x^j]= -\frac12 [x^i,F^j]+\frac12 [x^j,F^i]\\
=\frac12
(\theta^{jk}\frac{\partial F^i}{\partial x^k} - \theta^{ik}\frac{\partial F^j}{\partial x^k}) +\frac12( g^{jk}\frac{\partial F^i}{\partial\dot x^k}- g^{ik}\frac{\partial F^j}{\partial\dot x^k}),
\end{multline}
and $g^{ij}$ a symmetric matrix (this is a simple consequence of $\frac{D}{Dt}[x^i,x^j]=0)$. The first set of equations (\ref{h-a}) comes from the Jacobi identity $[x^i,[x^j,\dot x^k]] +(ijk)=0$. The equations (\ref{h-b}) come from the derivative of the first bracket in (\ref{NCR}) symmetrizing and adding the two equations.  The equations (\ref{h-c}) come from the derivative of the definition of the matrix $B$. The matrix $B$ in (\ref{B}) can be constructed from the equations (\ref{h-b}) and using the definition of the matrix $g^{ij}$.

We can write these conditions as the Helmholtz conditions plus terms that depend on the noncommutative parameter $\theta$
\begin{subequations}
\label{helmcond}
\begin{align}
&g_{ij}=g_{ji}, \label{hc-a}\\
&\frac{\partial g_{ij}}{\partial\dot{x}^k}-\frac{\partial g_{ik}}{\partial\dot{x}^j}+L^\theta_{ijk}=0, \label{hc-b}\\
&\frac{D}{Dt}g_{ij}=-\frac{1}{2}\left(g_{ik}\frac{\partial F^k}{\partial\dot{x}^j}+g_{jk}\frac{\partial F^k}{\partial\dot{x}^i}\right)+M^\theta_{ij}, \label{hc-c}\\
&\frac{D}{Dt}t_{ij}= g_{ik}\frac{\partial F^k}{\partial x^j}-g_{jk}\frac{\partial F^k}{\partial x^i}+N^\theta_{ij},\label{hc-d}
\end{align}
\end{subequations} 
where
$$t_{ij}=\frac12(g_{ik}\frac{\partial F^k}{\partial\dot{x}^j}-g_{jk}\frac{\partial F^k}{\partial\dot{x}^i}),$$
and $L^\theta_{ijk}, M^\theta_{ij}, N^\theta_{ij}$ are terms that depend on $\theta$. Explicitly  they  are given by
$$L^\theta_{ijk}=\theta^{sl}g_{sk}\frac{\partial g_{ij}}{\partial x^l}- \theta^{rl}g_{rj}\frac{\partial g_{ki}}{\partial x^l},$$
$$M^\theta_{ij}=-\frac12 g_{ir}(\theta^{rn}\frac{\partial F^{s}}{\partial x^n}+\theta^{sn}\frac{\partial F^{r}}{\partial x^n})g_{js},$$
$$N^\theta_{ij}=g_{li}g_{mj}(-\frac{D}{Dt}b^{lm}+b^{lk}\frac{\partial F^m}{\partial\dot{x}^k}-b^{mk}\frac{\partial F^l}{\partial\dot{x}^k})+ g^{lk}(t_{lj}M^\theta_{ik}-t_{li}M^\theta_{jk}),$$
where
$$b^{ij}=\frac12
(\theta^{jk}\frac{\partial F^i}{\partial x^k} - \theta^{ik}\frac{\partial F^j}{\partial x^k}).$$
Here $g_{ij}$ denotes the inverse matrix of our previous $g^{ij}$.
If one insist in the existence of a Lagrangian for the original system (\ref{eq-m}) or if it already exists then the subsidiary  conditions $L^\theta_{ijk}=0, M^\theta_{ij}=0, N^\theta_{ij}=0$ must be satisfied, namely
\begin{subequations}
\label{sub-cond}
\begin{align}
\theta^{in}\frac{\partial g^{jk}}{\partial x^n}- \theta^{jn}\frac{\partial g^{ik}}{\partial x^n}=0,\label{sc-a}\\
\theta^{in}\frac{\partial F^{j}}{\partial x^n}+\theta^{jn}\frac{\partial F^{i}}{\partial x^n}=0,\label{sc-b}\\
-\frac{D}{Dt}b^{lm}+b^{lk}\frac{\partial F^m}{\partial\dot{x}^k}-b^{mk}\frac{\partial F^l}{\partial\dot{x}^k}=0.\label{sc-c}
\end{align}
\end{subequations}
These equations impose very restrictive conditions on the forces $F^i$. As an example consider the case of a force proportional to $r^n$ where $r=\sqrt{\sum _i x^i x^i}$. The subsidiary conditions (\ref{sub-cond}) imply that the only solutions for $n$ are $n=0,1,2$. So, among the isotropic central forces the only system that admit both CCR and NCR quantization is the harmonic oscillator. This means that at least we can give two different pairs $(\sigma,H)$ which have the same second order equations of motion in configuration space.
In spite of the classical equivalence of the two formulations the resulting quantum theories are nonequivalent. This follows from the fact that the parameter $\theta$ breaks the degeneracy of the harmonic oscillator spectra. Explicitly for a two dimensional isotropic oscillator quantized using NCR we have
\begin{equation}
\label{eo}
E_{m,n}=\omega[(m+n)+\theta\omega(m-n)+1],
\end{equation}
where $n,m$ are two non-negative integers and $\omega$ is the frequency of the isotropic oscillator. For details of this calculation we refer the reader to the appendix. To fully appreciate the content of this quantization of the harmonic oscillator it is important to notice that here we are starting from the equations of motion of the harmonic oscillator and a compatible symplectic structure, {\em i.e.}, a solution of eqs. (\ref{helmcond}), given the force of the isotropic harmonic oscillator. The classical system defined in this way is the standard harmonic oscillator.
Upon quantization, using standard methods we arrive at the nonstandard result (\ref{eo}). The quantization of the noncommutative oscillator \cite{NP} is quite different. There the starting point is a NCCS with the Hamiltonian of the isotropic oscillator in two dimensions and a symplectic structure that implements the 
noncommutativity of the configuration space coordinates. This system is not the standard harmonic oscillator. Its quantization also produces a spectra that depends on the parameter $\theta$. In the next section we will study these NCCS as a special case of the results that we have obtained in this section.

\section{Generic solution of NCR dynamical consistency conditions}

To find  a solution of the dynamical consistency conditions (\ref{helm}) we will try to explore the type of forces that these conditions allow. The forces that come from a potential function are ruled out because the subsidiary conditions on them are too restrictive. We wonder if it is possible to guess a generic form of the forces that are compatible with the dynamical consistency conditions.
 Fortunately there exists at least one type of generic forces that allow us to find a solution to the consistency conditions. This solution is the analogous of the corresponding solution of the Helmholtz conditions for forces that are derivable from a potential function. Consider forces of the form
\begin{equation}
\label{nc-f}
F^i=-\frac{\partial V}{\partial x^i}+\theta^{ij}\frac{d}{dt}\frac{\partial V}{\partial x^j},
\end{equation}
for a ``potential function'' $V(x)$. This type of forces arise in the first order formulation for a Hamiltonian function of the form $H=\frac12 p^2+V$ with the symplectic structure
\begin{equation}
\label{sym-xp}
[x^i,x^j]=\theta^{ij},\quad [p_i,p_j]=0,\quad [x^i,p_j]=\delta^i_j,
\end{equation}
{\it i.e.}, from a NCCS with potential $V$.
The Hamilton equations of motion associated with a NCCS are
\begin{subequations}
\label{h-eq}
\begin{align}
\dot x^i + \theta^{ij}\dot{p_j}- p_i=0,\label{he-a}\\
-\dot{p_i} -\frac{\partial V}{\partial x^i}=0.\label{he-b}
\end{align}
\end{subequations}
Forces in configuration space of the form (\ref{nc-f}) can be deduced from these first order equations by using the set (\ref{he-b}) and taking the derivative with respect to time of the first set of equations (\ref{he-a}). Notice that this reduction of phase space variables to the configuration space is nonstandard, {\em i.e.}, the momenta are not auxiliary variables. So we can not use this reduction to obtain a Lagrangian for the system whose forces are of the form (\ref{nc-f}). This reduction is not ``variationally admissible'' \cite{Novikov}.   In particular the map from the space of initial conditions of the first order system (\ref{h-eq}),  $x^i(t_0)=x^i_0, p_j(t_0)={p_j}_0$, where $t_0$ is the initial time, to the configuration space is quite not standard. This map is
\begin{eqnarray}
x^i(t_0)=x_0^i,\\
\Big[ -\theta_{ik}\frac{\partial V}{\partial x^k}+
{\dot x}^i\Big] (t_0)  = {p_i}_0.
\end{eqnarray}
An interesting observation is that these initial conditions are not the usual initial position and velocities of Newtonian Mechanics but depend on the symplectic matrix $\sigma$ and the dynamics through the ``potential function'' $V(x)$. Of course we can recover the standard case when we use CCR.

To obtain the conditions on the integrating factor $g^{ij}$ and the function $V$, we will use the forces (\ref{nc-f}) in the dynamical consistency conditions (\ref{helm}). The result is
\begin{subequations}
\label{nc-helm}
\begin{align}
\theta^{il}\frac{\partial g^{il}}{\partial {x}^l}-\theta^{jl}\frac{\partial g^{ik}}{\partial x^l}=0,\label{nch-a}\\
\frac{d}{dt}g^{ij}=\frac{1}{2}\left(g^{ik}\theta^{jl}V_{kl}-\theta^{il}V_{jl}+\theta^{ik}\theta^{jl}\frac{d}{dt}V_{lk} + (i\leftrightarrow j)\right),\label{nch-b}\\
\frac{D}{Dt} B^{ij}=g^{il}V_{jl}-\theta^{jl}g^{ik}\frac{d}{dt}V_{lk}+\nonumber\\
\frac12 \theta^{jl}V_{lk}\left(\theta^{ik}V_{kr}-\theta^{kr}V_{ir}+\theta^{ir}g^{kn}V_{rn}-\theta^{kr}g^{in}V_{rn}\right)-(i\leftrightarrow j), \label{nch-c}
\end{align}
\end{subequations}
where
\begin{equation}
\label{nc-B}
B^{ij}=\frac12\left(\theta^{il}V_{jl}-\theta^{jl}g^{in}V_{ln}- \theta^{jl}V_{il}
+\theta^{il}g^{jn}V_{ln}\right),
\end{equation}
restricting the dependence of $g^{ij}$ to  the coordinates and using the notation  $V_l\equiv\frac{\partial V}{\partial x^l}$. 

From equation (\ref{nch-b}) we can guess a solution of the form
\begin{equation}
\label{sol-g}
g^{ij}=\delta^{ij}+\theta^{in}\theta^{jm}V_{nm}.
\end{equation}
It turns out that this guess is in fact correct. It solves all the remanning conditions {\em for any function} $V(x)$. We conclude that the symplectic form
\begin{subequations}
\label{nc-NCR}
\begin{align}
[x^i,\dot x^j]=\delta^i_j+\theta^{in}\theta^{jm}V_{nm},\quad [x^i,x^j]=\theta^{ij},\label{nc-NCR-a}\\
[\dot x^i,\dot x^j]=\theta^{il}V_{jl}+\frac12\theta^{il}\theta^{jr}\theta^{ns}V_{ln} V_{rs}-(i\leftrightarrow j),\label{nc-NCR-b}
\end{align}
\end{subequations}
is compatible with the system
\begin{equation}
\label{nc-eq-m}
\ddot x^i+\frac{\partial V}{\partial x^i}-\theta^{ij}\frac{d}{dt}\frac{\partial V}{\partial x^j}=0. 
\end{equation}
This imply that the classical system defined by the equations of motion (\ref{nc-eq-m}) and the symplectic form (\ref{nc-NCR})
can be consistently  quantized.

Two comments are in order. The relation between the momenta and the velocities given in (\ref{h-eq}) 
\begin{equation}
\label{nc-p}
p_i=\dot x^i - \theta^{ij}\frac{\partial V}{\partial x^j},
\end{equation}
with the identity mapping for the configuration space can be used to map the symplectic form (\ref{nc-NCR}) to the symplectic form (\ref{sym-xp}). So the equations of motion (\ref{nc-eq-m}) can be obtained  from the pair $(\sigma, H)$ as we have presented them in the previous paragraph or it can be deduced from (\ref{nc-NCR}) with Hamiltonian
\begin{equation}
\label{nc-h}
H=\frac12 (\dot x^i - \theta^{ij}\frac{\partial V}{\partial x^j})^2 +V.
\end{equation}
The two formulations for the system (\ref{nc-eq-m}) are classically equivalent but, in general, they give different quantum theories. In this way we have two classically equivalent formulations of the noncommutative dynamics given by a general Hamiltonian function $H=T+V$ whose associated quantum theories are different. If we quantize the noncommutative harmonic oscillator of \cite{NP} using (\ref{nc-h}) and (\ref{nc-NCR}) we will obtain a different energy spectra with a different dependence on the parameter $\theta$.

Notice the important fact that in general (up to the linear and quadratic potentials) the system does not have a Lagrangian, {\em i.e.}, the solution presented in (\ref{nc-NCR}) is not a solution for the Helmholtz conditions.
Nevertheless it can be quantized in a consistent way. 

As an example of a NCCS that does not admit a variational formulation we can consider the system defined by the potential function $V=\frac12 \omega^2 x^2$ in two dimensions. It is easy to prove that this simple system does not have a second order variational formulation but nevertheless it can be consistently quantized. The system is
\begin{align}
\ddot{x}&=-\omega^2 x, \\
\ddot{y}&=-\theta\omega^2 \dot{x}.
\end{align}
Using the first order representation given by the standard Hamiltonian $H=\frac12 p^2+V$ with the associated momenta given by (\ref{nc-p}),
the symplectic form is
$$[x,y]=\theta,\quad  [x,p_x]=1,\quad [y,p_y]=1.
$$
The other brackets are equal to zero. Taking as a basis for the Hilbert space $|x',p'_y\rangle$ the propagator is
\begin{multline}
\label{resultado}
\langle x'',p''_y;t''|x',p'_y;t'\rangle=\frac{e^{D}}{\sqrt{\sin(\omega T)}}e^{-\frac{i}{2}k^2 T+i\frac{\omega}{2}\cot(\omega T)({x'}^2+{x''}^2)-i\frac{\omega}{\sin(\omega T)}x'x''},
\end{multline}
where $k$ is the constant value of $p_y$, and $T=t''-t'$. The constant $D$ can be determined using the condition that when we take the limit of (\ref{resultado}) as $t''\rightarrow t'$ the propagator is a Dirac delta. This result does not depends on the noncommutative parameter $\theta$ but this is an effect of the choice of the basis for the Hilbert space.

Our second comment is concerned with the construction of the bracket of any two variables at different times. Using the notation $z^\alpha=(x^i,\dot x^i),\,\, \dot z^\alpha=(\dot x^i, F^i),\,\, \ddot z^\alpha=(F^i,\frac{DF^i}{Dt})....$ where $\alpha=1,2,...2n$ and  $i,j=1,2,...n$, the Poisson brackets at different times are
$$[z^\alpha(t),z^\beta(t+\delta t)]=\sigma^{\alpha\beta}+{\cal G^{\alpha\beta}}(t)\delta t +\frac12\left(\frac{D}{Dt}{\cal G^{\alpha\beta}}+
{\cal B}^{\alpha\beta}\right)(\delta t)^2+\ldots,$$
up to second order in $\delta t$ where
$${\cal B}^{\alpha\beta}=[\dot z^\alpha(t),\dot z^\beta(t)],\qquad {\cal G}^{\alpha\beta}=[z^\alpha(t),\dot z^\beta(t)],$$
and 
$$\sigma^{\alpha\beta}=\left(\begin{array}{cc}\theta^{ij} & g^{ij} \\-g^{ij} & 
B^{ij}\end{array}\right),$$
with $g^{ij}$ and $B^{ij}$ defined in the usual way. The relation between this brackets and the Peierls brackets is outside the scope of the present letter. We only mention that in the limit $\theta\to 0$ we can relate this result with the standard result obtained from the covariant Peierls brackets \cite{deWitt}.
In that case we can construct from the Green function the CCR at different times. When a Lagrangian for the equations of motion exists the result is \cite{deWitt}
\\
\\
for $\alpha=i,\beta=j$
$$
G^{ij}(t,t+\delta t)=  g^{ij}\delta t +\frac12\left(\frac{D g^{ij}}{Dt}+B^{ij}\right)(\delta t)^2+\ldots,$$
for $\alpha=i,\beta=j+n$
$$
G^{ij}(t,t+\delta t)= g^{ij} +\left(\frac{D g^{ij}}{Dt}+B^{ij}\right)(\delta t)+\ldots,$$
for $\alpha=i+n,\beta=j+n$
$$
G^{ij}(t,t+\delta t)= B^{ij} +O(\delta t)+\ldots,$$
where in terms of a second order Lagrangian
$$g_{ij}=\frac{\partial L}{\partial\dot x^i\partial\dot x^j},$$
$$B^{ij}=g^{ik}\left(\frac{\partial L}{\partial x^k\partial\dot x^l}-\frac{\partial L}{\partial\dot x^k\partial x^l}\right) g^{lj}.$$
The equal time Poisson brackets are
$$[x^i,x^j]=0,\quad [x^i,\dot x^j]=g^{ij},\quad [\dot x^i,\dot x^j]=B^{ij},$$
as expected.

\section{Conclusions}

In this letter we have studied the problem of the dynamical compatibility  between a given set of equations of motion in configuration space and a symplectic structure.
Generalizing a previous result that states that this dynamical compatibility can be related to the inverse problem of the calculus of variations we have presented here new dynamical conditions that could be seen as a deformation of the Helmholtz conditions.  We were limiting ourselves to the case where the bracket between the configuration variables is a constant antisymmetric matrix. 
Surprisingly the deformed problem has a generic solution for the new dynamical compatibility conditions in terms of a ``potential function'' for the forces of the form (\ref{nc-f}). 

We can ask if these new conditions are the integrability equations of the inverse problem of the calculus of variations of some sort of generalized noncommutative dynamics. It could be the case that a particular deformation of the equations of motion by adding some $\theta$ dependent terms can be related with the conditions obtained here. This is an open interesting problem that we leave for a future work.   

We have also presented the deformed  brackets at different times and we have compared them with the well known Peierls brackets associated with CCR. Our brackets at different times are the Peierls brackets when the deformation parameter is zero. The question if it is possible to relate the deformed brackets at different times with the Green function associated to the equations of motion is outside the scope of the present letter.  

Finally we will comment on the extension of our results to the case of field theory. The symplectic noncommutativity studied here have important consequences for the formulation of the corresponding field theory. In addition to the problem of the existence of a Lagrangian, the problem of the construction of a meaningful field theory with a well defined perturbative framework, a vacuum and asymptotic states is not yet solved. Some drastic changes to the standard formulation of field theoretic methods are needed to solve these relevant questions. Nevertheless we can speculate on some of the consequences that nonequivalent quantization can produce in field theory. It seems  that the quantization of a Klein-Gordon equation for a set of non-interacting scalar fields using NCR produces a quantum interacting field theory that can be based on the oscillator quantization presented in the appendix. From the other hand it is not difficult to see that a Lagrangian function for a theory like $\phi^4$ in the NCR approach does not exists.  
 
It is also worth noticing that by extending our results of section 3 to field theory we could produce models that have some contact with Double Special Relativity (DSR) \cite{MS}. We can see that in fact this could be the case because an equation of motion of the form (\ref{nc-eq-m}) will produce in field theory some sort of deformed dispersion relations that in simple cases have the form of the deformed dispersion relations presented in recent DSR literature.

\section*{Acknowledgments} 

This work was supported in part by grants CONACyT 32431-E and DGAPA IN104503. A. G. wishes to thanks to S. Hojman for a stimulating and useful discussion.

\section*{Appendix}

Let us consider the system of equations of motion that describe the isotropic harmonic oscillator in two dimensions 
\begin{equation}
\ddot{x}^i=-\omega^2 x^i.
\end{equation}
According to our previous analysis this system admits at least two different consistent quantizations: a) the standard quantization, b) a NCR quantization. We will present here the b) quantization scheme. The symplectic structure is
$$
[x,y]=\theta, \qquad [x,\dot{x}]=[y,\dot{y}]=1, \qquad [\dot{x},\dot{y}]=\theta\omega^2,
$$
and the other brackets are zero.

Defining a first order system with the momenta given by
$$
p_x=\dot{x}, \qquad p_y=\dot{y},
$$
the Hamiltonian can be constructed and is
$$
H=\frac{1}{1-\theta^2\omega^2}\left[\frac{1}{2}(p_x^2+p_y^2)+\frac{\omega^2}{2}(x^2+y^2)+\theta\omega^2(x p_y-y p_x)\right]. 
$$
We choose a basis for the Hilbert space $\{\mid x,p_y\rangle\}$ such that the operators $\hat{x}$ and $\hat{p}_y$ acts in a multiplicative way, and 
\begin{align}
\label{PYrepre1}
&\langle x,p_y\mid\hat{p}_x=[-i\frac{\partial}{\partial x}+i\theta\omega^2\frac{\partial}{\partial p_y}]\langle x,p_y\mid ,\\\label{PYrepre2}
&\langle x,p_y\mid\hat{y}=[i\frac{\partial}{\partial p_y}-i\theta\frac{\partial}{\partial x}]\langle x,p_y\mid.
\end{align}

The Hamiltonian can be expressed in terms of the energy operator $\hat{E}=\frac{1}{2}(\hat{p}_x^2+\hat{p}_y^2)+\frac{\omega^2}{2}(\hat{x}^2+\hat{y}^2)$ and the angular momentum operator $\hat{L}=\hat{x}\hat{p}_y-\hat{y}\hat{p}_x$, which commute with each other. An energy and angular momentum basis can be constructed using creation and annihilation operators of ``+" and ``-" quanta \cite{Messia}. 

Given the usual $a_1=\frac{1}{\sqrt{2\omega}}(\omega\hat{x}+i\hat{p}_x)$ and $a_2=\frac{1}{\sqrt{2\omega}}(\omega\hat{y}+i\hat{p}_y)$, define the operators 
$$
A_{+}=\frac{1}{\sqrt{2}}(a_1-ia_2), \qquad A_{-}=\frac{1}{\sqrt{2}}(a_1+ia_2),
$$
and then
$$
N_{\pm}=A_{\pm}^{\dagger}A_{\pm},
$$
in terms of which the energy and angular momentum reads
$$
\hat{E}=\omega(N_{+}+N_{-}+1),
$$
$$
\hat{L}=N_{+}-N_{-}.
$$
The operators $A_{\pm}$ and $A_{\pm}^{\dagger}$ satisfies the commutators 
$$
[A_{\pm},A_{\pm}^{\dagger}]=(1\mp\theta\omega),
$$
and the rest are zero. So, the commutator of these operators with $N_{+}+N_{-}$ and $\hat{L}$ gives
$$
[N_{+}+N_{-},A_{\pm}]=-(1\mp\theta\omega)A_{\pm}, \quad [N_{+}+N_{-},A_{\pm}^{\dagger}]=(1\mp\theta\omega)A_{\pm}^{\dagger}, 
$$
$$
[\hat{L},A_{\pm}]=\mp(1\mp\theta\omega)A_{\pm}, \quad [\hat{L},A_{\pm}^{\dagger}]=\pm(1\mp\theta\omega)A_{\pm}^{\dagger}.
$$
With this information we can conclude that $A_{\pm}$ and $A_{\pm}^{\dagger}$ annihilates and creates quanta. If $\mid\eta,\lambda\rangle$ is an eigenvector of $N_{+}+N_{-}$ and $\hat{L}$ with eigenvalues $\eta$ and $\lambda$ respectively, we then have
$$
A_{\pm}\mid\eta,\lambda\rangle\sim\mid\eta-(1\mp\theta\omega),\lambda\mp(1\mp\theta\omega)\rangle,
$$
and
$$
A_{\pm}^{\dagger}\mid\eta,\lambda\rangle\sim\mid\eta+(1\mp\theta\omega),\lambda\pm(1\mp\theta\omega)\rangle.
$$
From the positivity of the energy we know that a state that is annihilated by $A_{\pm}$ exists and can be written as
$$
A_{\pm}\mid0,\lambda_{0}\rangle=0\Rightarrow\langle x,p_y\mid A_{\pm}\mid0,\lambda_{0}\rangle=0.
$$
From this two equations and using the representation given by (\ref{PYrepre1}) and (\ref{PYrepre2}) we found  the normalized wave function for the lowest energy state 
$$
\psi_{0,\lambda_{0}}(x,p_y)=\frac{1}{(1-\theta^2\omega^2)^\frac{1}{4}\sqrt{\pi}}e^{-\frac{1}{1-\theta^2\omega^2}(\frac{\omega}{2}x^2+\frac{1}{2\omega}p_y^2+\theta\omega xp_y)}.
$$
Now, from $\hat{L}\mid0,\lambda_{0}\rangle=\lambda_{0}\mid0,\lambda_{0}\rangle$, we can calculate $\lambda_{0}=-\theta\omega$.

From this vacuum state the energy spectrum can be calculated. The result is
$$
\mid(m+n)+\theta\omega(m-n),-(m-n)-\theta\omega(m+n+1)\rangle=
$$
$$\frac{1}{2\sqrt{m!n!}}(A_{+}^{\dagger})^{n}(A_{-}^{\dagger})^{m}\mid0,-\theta\omega\rangle,
$$
with the energy spectrum  given by
$$E_{m,n}=\omega[(m+n)+\theta\omega(m-n)+1].$$

\end{document}